\documentclass{article}

\usepackage{arxiv}

\usepackage[utf8]{inputenc} 
\usepackage[T1]{fontenc}    
\usepackage{url}            
\usepackage{booktabs}       
\usepackage{amsfonts}       
\usepackage{nicefrac}       
\usepackage{microtype}      
\usepackage{lipsum}

\usepackage[english]{babel}
\usepackage{float}
\usepackage{amsmath}
\usepackage{amsfonts}
\usepackage{graphicx}
\usepackage{algorithm}
\usepackage{algpseudocode}
\usepackage[colorinlistoftodos]{todonotes}
\usepackage[colorlinks=true, allcolors=blue]{hyperref}

\title{CS Sparse K-means: An Algorithm for Cluster-Specific Feature Selection in High-Dimensional Clustering}

\author{
  Xiangrui Zeng \\
  Computational Biology Department\\
  Carnegie Mellon University\\
  Pittsburgh, PA 15213 \\
  \texttt{xiangruz@andrew.cmu.edu} \\
   \And
  Hongyu Zheng \\
  Computational Biology Department\\
  Carnegie Mellon University\\
  Pittsburgh, PA 15213 \\
  \texttt{hongyuz1@andrew.cmu.edu} \\
}

\begin{document}

\maketitle

\begin{abstract}
Feature selection is an important and challenging task in high dimensional clustering. For example, in  genomics, there may only be a small number of genes that are differentially expressed, which are informative to the overall clustering structure.  Existing feature selection methods, such as Sparse K-means, rarely tackle the problem of accounting features that can only separate a subset of clusters. In genomics, it is highly likely that a gene can only define one subtype against all the other subtypes or distinguish a pair of subtypes but not others. In this paper, we propose a K-means based clustering algorithm that discovers informative features as well as which cluster pairs are separable by each selected features. The method is essentially an EM algorithm, in which we introduce lasso-type constraints on each cluster pair in the M step, and make the E step possible by maximizing the raw cross-cluster distance instead of minimizing the intra-cluster distance. The results were demonstrated on simulated data and a leukemia gene expression dataset.  
\end{abstract}


\section{Introduction}

Dimensionality reduction has been a perpetual
problem in machine learning and computational biology, especially in computational genomics when a typical set of gene expression data has a huge number of genes. Typical dimensionality reduction methods like PCA would work on such cases, resulting in a low dimensional space embedding, but are less biologically meaningful, since each feature vector is a weight map over all genes, so all genes contribute to each feature vector. 

In the field of computational genomics, for both classification and clustering problems, there usually exists a small set of so-called "signature genes" that are actually reverent to the problem. And that is why the more common approach for dimensionality reduction in the field is to perform feature selection, that is, to select a number of features (genes in our case), and perform classification or clustering only with the inferred set of genes (or in a more flexible manner, weight important genes higher and give zero weight to genes that are similarly expressed between clusters). For applications of feature selection in computational biology, see \cite{appl1, appl2, appl3, appl4}.

In this study, we focus on clustering problems. Here, the cluster structure can represent subtypes of samples, and as described before, data points (samples) might only differ in a small subset of features (genes), which are important in the differentiation of the subtypes. Finding the important genes that distinguish subtypes also helps the interpretation of the disease states because selected genes can be used for further pathway and network analysis. 

\begin{figure}[H]
\centering
\includegraphics[width=0.8\textwidth]{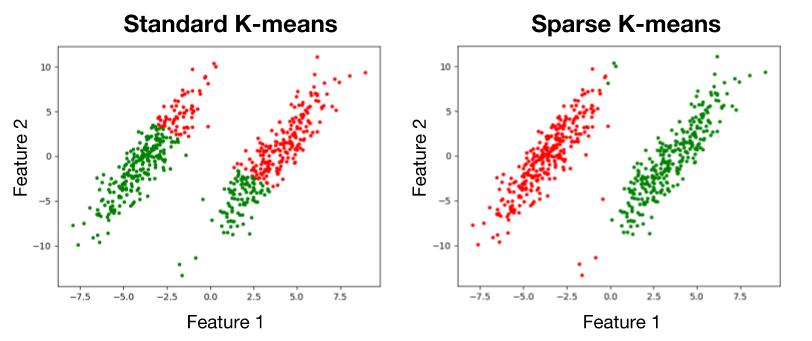}
\caption{In a 2D example, two clusters only differ on feature 1. Sparse 2-means gives better results because it only considers feature 1. Feature 1's weight is close to 1 and feature 2's weight is close to 0.}
\label{fig:2means}
\end{figure}

Sparse K-means \cite{witten2010framework} is an algorithm that performs two tasks: it assigns a weight to each feature (gene) between 0 and 1, and clusters on high-dimensional data. These two steps are dependent on each other; the weight is calculated with clustering results, and the clustering is based on feature weights, and an Expectation-Maximization couples the tasks and results in both a good clustering that takes feature sparsity into account, and a good weight assignment that is based on the clustering. An illustration of Sparse K-means can be found in Figure \ref{fig:2means}.


There is one shortcoming of the aforementioned method. It assigns an importance score (weight) to a feature across all clusters, which is intuitive since in feature selection frameworks, the importance score is designed to be the importance of a feature to the whole cluster structure. Nevertheless, it is common for a gene to contribute to a particular subtype but not others, and the Sparse K-means does not take this into consideration. Ability to infer the importance for each feature between clusters can also be beneficial in practice, since knowing the subtype-specific gene importance helps the interpretation and diagnostics of different subtypes.

\begin{figure}[H]
\centering
\includegraphics[width=0.9\textwidth]{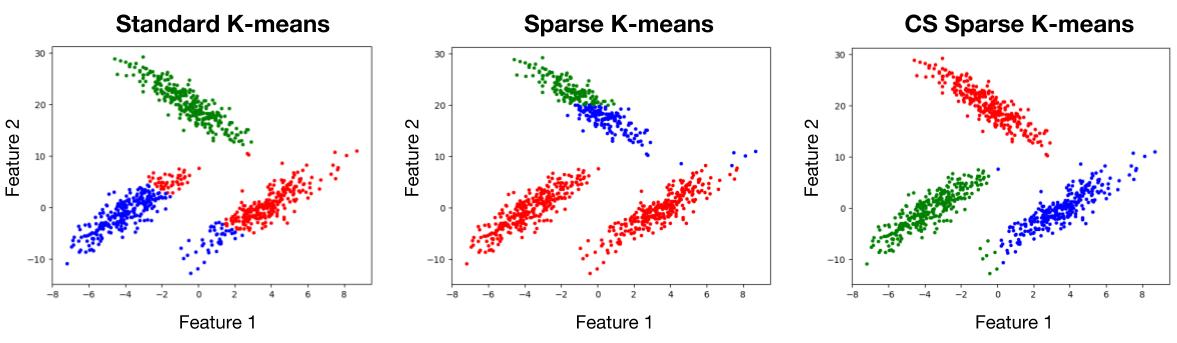}
\caption{In a 2D example, the cluster pair 1 (green in the right figure) \& 2 (blue) differs only on feature 1; cluster pair 1 \& 3 (red) and 2 \& 3 differ mainly on feature 2.  CS Sparse 3-means gives better results because it assigns corresponding weight both on features and cluster pairs. Sparse 3-means gives bad results because only feature 2 is considered.}
\label{fig:3means}
\end{figure}

Therefore, We propose CS sparse K-means (CS stands for Cluster-Specific), which clusters the samples using not only the adaptively selected features but also their contribution to each pair of clusters. The idea of CS sparse K-means is illustrated in Figure \ref{fig:3means}. The weight assignment of CS sparse K-means is shown in Table \ref{tab:we}.


\begin{table}[H]
\label{tab:we}
\begin{center}
	\begin{tabular}{  c | c  c  c }
    \hline
    Weight & Cluster 1 vs 2 & Cluster 1 vs 3 & Cluster 2 vs 3 \\ \hline 
    Feature 1 & 0.895 & 0.030 & 0.050  \\ 
    Feature 2 & 0.447 & 0.999 & 0.999  \\ \hline
    \end{tabular}
    \vspace{1em}
    \caption{Weight assignment of features by cluster pairs. Clearly, feature 1 is mainly responsible in distinguishing cluster 1 \& 2, and feature 2 is mainly responsible for distinguishing cluster 1 \& 3 and cluster 2 \& 3.}
\end{center}
\end{table}

\section{Related Work}

Many feature selection methods have been developed \cite{witten2010framework, survey2014}. In general, there are two ways to perform feature selection, namely filter approach and wrapper approach; The former approach calculates some relevance score on features and discards those with low scores. The latter approach uses a predictor as a black box and aims to obtain a subset of features that maximize the predictor score over the said subset. Wrapper methods are more common in practice due to more control over the selection process and direct association with final results, and in this paper we will mostly focus on such algorithms.

Our proposed method can be seen as an extension of classical Sparse K-Means algorithm, but it's worth noting there are other works that also focus on the same problem, but with a much different approach. An earlier attempt comes from the paper titled "Pairwise Variable Selection for High-Dimensional Model-Based Clustering \cite{guo2010pairwise}". They introduce the following regularization term to the traditional Gaussian Mixture models($k$ over all features, $i < j$ over all cluster pairs, $\lambda$ is a tunable hyperparameter):

$$
\lambda \sum_{k} \sum_{i < j} |\mu_{ik} - \mu_{jk}|
$$

Since L1-norm enforces sparsity, if for a feature $k$ and two clusters $i < j$ we have $\mu_{ik} = \mu_{jk}$, then this feature is considered not useful in separating cluster $i$ and $j$.

The authors develop an EM algorithm to iteratively update means and probabilistic assignments.

We note that the method is simple and intuitive, but lacks intuitive control over desired sparsity and only gives binary results(the feature is useful or is not useful instead of assigning a value for the situation).

The paper titled "CRAFT: ClusteR-specific Assorted Feature selecTion" \cite{garg2016craft} uses Bayesian method for this problem. The three main parts of their models are:

1. Whether a feature should be selected for a cluster is indicated by binary variable $v_{kd}$ which follows a Bernoulli distribution and the parameter for this Bernoulli is drawn from a Gaussian with predefined mean and variance.

2. The cluster assignments and the number of clusters are generated by a Dirichlet Process.

3. If a feature is selected for a cluster, its distribution within the cluster follows $N(\mu_{kd}, \sigma_{kd})$, and if not, it follows $N(\mu_d, \sigma_d)$ (independent of clusters).

The authors then developed approximate inference algorithms for the parameters.

However, the process has a lot of parameters and requires several regularizations to work, while our methods are much simpler to interpret. Its upsides include being able to handle categorical features (not mentioned above) and inference over the number of clusters by virtue of Dirichlet Process.

\section{Methods}

\subsection{Setup}

Suppose we have a set of $n$ samples and $k$ features that take on real values (Denoted $X\in \mathbb{R}^{n\times k}$), and a predefined number of clusters $C$. The classical clustering task asks for a partition $z:[n]\rightarrow [C]$ that minimizes the inter-cluster distance:
$$
\min	\sum_{k\in[C]}\frac{1}{2|C_{k}|}\sum_{i,j\in C_{k}}||X_{i}-X_{j}||^{2}
$$

We are however inspired by the so-called un-normalized inter-cluster distance, which only differs by a scaling constant if the clusters are of similar size ($C_i \approx C_j$):

$$
\min	\frac{1}{2}\sum_{k\in[C]}\sum_{i,j\in C_{k}}||X_{i}-X_{j}||^{2}
$$

For a fixed dataset, the quantity $\sum_{i,j} ||X_{i}-X_{j}||^{2}$ is the same regardless of the partitions. This means the above optimization problem is equivalent to the following, which we call un-normalized cross-cluster distance:

$$
\max	\frac{1}{2} \sum_{i,j, z(i)\neq z(j)}||X_{i}-X_{j}||^{2}
$$

We now introduce the parameters of our algorithm. $w^{(k)}_{i,j}$ is a scalar that defines importance of feature $k$ in separating cluster $i$ and $j$. Given the above formulation, it's natural to scale the result based on the parameters we just introduced:

$$
\max \frac{1}{2} \sum_{k} \sum_{i,j, z(i)\neq z(j)} w^{(k)}_{z(i),z(j)} (X_{ik}-X_{jk})^2
$$

And this will be the target function in our algorithm. Fix dataset $X$, we can write the function as $F(w, z) = \frac{1}{2} \sum_{k} \sum_{i,j, z(i)\neq z(j)} w^{(k)}_{z(i),z(j)} (X_{ik}-X_{jk})^2$; We next introduce constraints on $w$ and design an E-M algorithm to perform clustering.

\subsection{Expectation-Maximization}

In the E-step, we fix $w$ and find an assignment $z$ that will maximize $F(w,z)$. As the title suggests, we are going to use K-Means at this step to get an assignment. However, since K-Means is not optimal, there is no formal guarantee(also as we mentioned above, K-Means was not designed for this objective, but they are similar if cluster sizes are close); Rather, we will later show that this strategy works well in practice.

First, note that we defined a distance metric between any two points $a, b\in \mathbb{R}^{k}$, only if their cluster assignment is given as $z(a)$ and $z(b)$:

$$
D(a, b) = \sum_{k} w^{(k)}_{z(a), z(b)} (a_k - b_k)^2
$$

However, in K-Means, the only distance we will ever compute is the distance between a data point and current mean of a cluster(defined as $\mu_c\in \mathbb{R}^{k}$ for each cluster $c$), which (at the time of computation) are assumed to be in the same cluster, but we did not define $D$ between two points in the same cluster. To circumvent this problem, we again take the idea of complementary: In original K-means, when assigning points, the rule is:

$$
z(i) = \text{argmin}_{c} ||X_i - \mu_c||^2
$$

This is actually equivalent to the following, since $\sum_c ||X_i-\mu_c||^2$ is a constant:

$$
z(i) = \text{argmax}_{c} \sum_{c'\neq c} ||X_i - \mu_{c'}||^2
$$

We can now naturally extend this idea for our case:

$$
z(i) = \text{argmax}_{c} \sum_{c'\neq c} \sum_k w^{(k)}_{c, c'} (X_{ik}-\mu_{c'k})^2
$$

In the M-step, we fix assignment $z$ and find a set of $w$ that will maximize $F(w,z)$. First of all, $F$ is a linear function of $w$:

$$
F(w,z) = \sum_k \sum_{a < b} w^{(k)}_{a, b} (\sum_{i\in C_a} \sum_{j \in C_b} (X_{ik} - X_{jk})^2) = \sum_{k, a<b} w^{(k)}_{a, b} s^{(k)}_{a, b}
$$

where $s$ are independent of $w$ and we know $s\geq 0$. Note that we can calculate $s$ with sufficient statistics without iterating over all pairs in $(C_a, C_b)$: $s^{(k)}_{a,b} = |C_b|(\sum_{i\in C_a}X_{ik}^2) + |C_a|(\sum_{j\in C_b} X_{jk}^2) - 2(\sum_{i\in C_a}X_{ik})(\sum_{j\in C_b} X_{jk})$.

Without constraint on $w$ we will just take it to infinity and it's not interesting. Inspired by Sparse K-Means, and Lasso in general, we introduce the constraint on $w$:

\begin{align*}
\sum_k w^{(k)}_{a, b} \leq T, &&\forall a < b \\
\sum_k (w^{(k)}_{a, b})^2 \leq 1, &&\forall a < b \\
w \geq 0 &&
\end{align*}

where $T$ is a tunable parameter. Note that this can be broken down to $c(c-1)/2$ instances of following optimization problem:

\begin{align*}
\max &&  w^Tx \\
\text{s.t.} && ||w||_1 \leq T \\
&& ||w||_2 \leq 1 \\
&& w \geq 0
\end{align*}

And the solution to this particular optimization problem is given by normalizing a parameterized soft-thresholding function:
\begin{align*}
w & =  \frac{S(x, \delta)}{||S(x, \delta)||_2} \\
S(x, \delta)_i & = \max (x_i - \delta, 0)
\end{align*}

The parameter $\delta$ is taken to be minimum non-negative value that satisfies $||w||_1 \leq T$(note that $||w||_2 = 1$ is guaranteed). There is no compact representation for the optimal $\delta$, but binary search would suffice.

As in general E-M algorithm, we alternative between E-step (updating $z$) and M-step (updating $w$) until the function $F(w,z)$ converges, or maximum iteration reached. The next section gives a formal description of the algorithm.

\subsection{Implementation}
We give a full description of our purposed Cluster-Specific Sparse K-Means on next page. We provide pseudo code with specifications of input and output, and both E and M steps necessary to perform the optimization. 

We don't explicitly define $F(w,z)$ at line 22, since in practice, the M-step function also returns sum of $s\cdot f(\delta)$ which is $F(w,z)$.

The algorithm has been implemented in Python 3. In practice, we set the convergence threshold to 0.001 as the target value is rather huge in most cases (analogous to total cross-cluster distance, so the value scales in a quadratic way with $n$). In the binary search process, the starting interval is set to $[0, \max(s) - \epsilon]$ to ensure at least one entry in $f(\delta)$ is positive.

\begin{algorithm}
\caption{E-Step, M-Step and main CSSKM process}
\hspace*{\algorithmicindent} 
\begin{tabular}{rl}
\textbf{Input} & $X$, observations \\
& $C$, number of clusters \\
& $T$, tunable parameter in M-step \\
\textbf{Output} & $w^{(k)}_{a,b}$, importance of feature $k$ between cluster $a$ and $b$ \\
& $z(i)$, assignment of data points to clusters
\end{tabular} \\ \\
\begin{algorithmic}[1]
\Function{E-step}{$X$, $C$, $w$} \Comment{Assigns observations based on importance value}
\State Initialize cluster centers $\mu_i$
\While {$\mu_i$ doesn't converge}
\State Assign $z(i) = \text{argmax}_{c} \sum_{c'\neq c} \sum_k w^{(k)}_{c, c'} (X_{ik}-\mu_{c'k})^2$ for each data
\State Recalculate centers $\mu_i = \frac{1}{|C_i|} \sum_{j\in C_i} X_j$
\EndWhile
\State \Return Latest $z(i)$
\EndFunction
\Function{M-step}{$X$, $C$, $T$, $z$} \Comment{Calculates importance value based on cluster assignment}
\For {each pairs of cluster $a < b$}
\State Calculate $s_k = \sum_{i\in C_a} \sum_{j \in C_b} (X_{ik} - X_{jk})^2$ using sufficient statistics
\State Define $f(\delta) = N(S(s, \delta))$ where $N(x) = x / ||x||_2$ and $S(s, \delta)_i = \max(s_i - \delta, 0)$
\State Binary search for smallest nonnegative $\delta$ such that $||f(\delta)||_1 \leq T$
\State Assign $w^{(k)}_{a,b} = f(\delta)_k$ for each feature $k$
\EndFor
\State \Return Completed $w$
\EndFunction

\Function{CSSKM}{$X$, $C$, $T$} \Comment{Main Process}
\State Pick $C$ random datapoints as centers of clusters
\State Set $z(i)$ to closest cluster center in normal L2 distance
\State $w$ = \Call{M-step}{$X$, $C$, $T$, $z$}
\While {$F(w,z)$ doesn't converge}
\State $z$ = \Call{E-step}{$X$, $C$, $w$}
\State $w$ = \Call{M-step}{$X$, $C$, $T$, $z$}
\EndWhile
\State \Return $z$, $w$
\EndFunction
\end{algorithmic}
\end{algorithm}

\newpage
\section{Results}
\subsection{Analysis of a simulated dataset}
We simulated a dataset of 60 samples and 100 features. The background followed $N(0,1)$. A small signal of 1.2 was added to feature 1 to 10 of sample 1 to 20, feature 11 to 20 of sample 21 to 40,  and feature 21 to 30 of sample 41 to 60. Therefore, we formed 3 clusters, each with 10 distinct features.

\begin{figure}[H]
\centering
\includegraphics[width=0.7\textwidth]{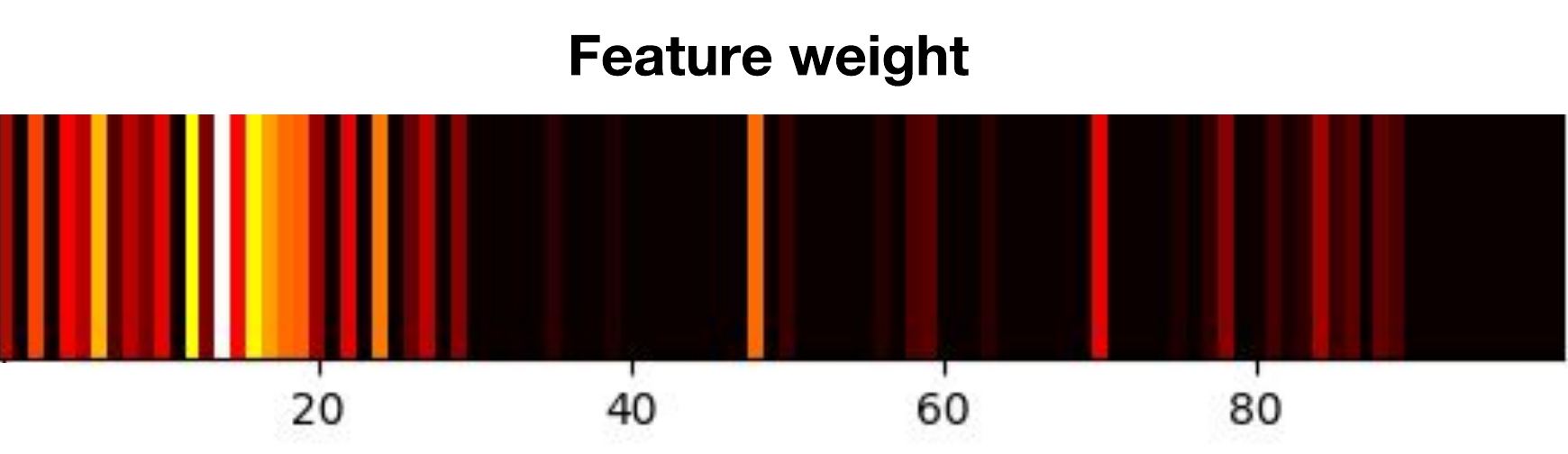}
\caption{In a simulated high-dimensional example, sparse 3-means assigned weight to each feature. It is clear that feature 1 to 30 are important to the overall cluster structure, although there are still some noise detected among feature 31 to 100.}
\label{fig:R-weight}
\end{figure}

Sparse 3-means and CS sparse 3-means were performed with $T = 5$. Both algorithms obtained relatively high accuracy: Our algorithm misclassified one sample (accuracy 98.3\%) while sparse K-means misclassified three (accuracy 95\%). The assigned weight for each feature by sparse K-means is shown in heatmap (Figure \ref{fig:R-weight}).

However, our algorithm is able to generate cluster(pair) specific weights:

\begin{figure}[H]
\centering
\includegraphics[width=0.8\textwidth]{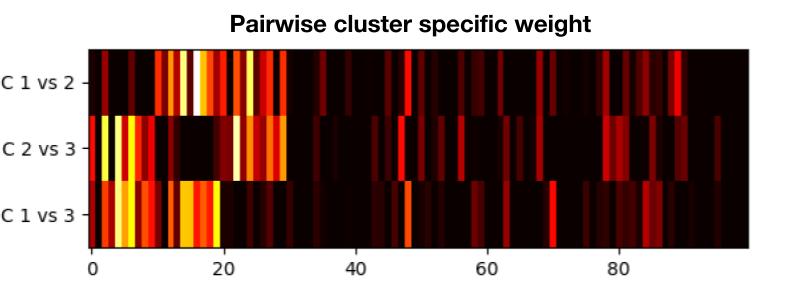}
\caption{In a simulated high-dimensional example, CS sparse 3-means assigned feature weight to each cluster pairs. It is clear that feature 1 to 10 define cluster 3 (cannot distinguish C1 vs C2 but can distinguish C1 vs C3 and C2 vs C3). Feature 11 to 20 define cluster 1 and feature 21 to 30 cluster 2.}
\label{fig:R-weight}
\end{figure}

Therefore, CS sparse K-means achieved better accuracy as well as discovered informative features specific to each cluster pair.

\subsection{Analysis of a leukemia dataset with three subtypes}
To demonstrate the efficacy of CS sparse K-means on a real-world dataset, we compared the performance of sparse 3-means and CS sparse 3-means on a leukemia gene expression dataset of 3 subtypes \cite{verhaak2009prediction}.

The preprocessed dataset \cite{huo2016meta} consists 89 samples and 5135 genes. The 89 bone marrow aspirates or peripheral blood samples are of 3 subtypes: 33 cohorts with ``inv(16)" abnormalities,  21 cohorts ``t(15;17)" abnormalities, and 35 cohorts with ``t(8;21)" abnormalities.

We applied sparse 3-means and CS sparse 3-means with T = 5 to this gene expression dataset. The sparse 3-means misclassified 4 samples (accuracy 95.5\%): two ``t(8;21)" abnormalities were misclassified as ``inv(16)" abnormalities, one ``t(8;21)" abnormality was misclassified as ``t(15;17)" abnormality, and one `inv(16)" abnormality was misclassified as ``t(8;21)" abnormality. The CS sparse 3-means misclassified 3 samples (accuracy 96.6\%): two ``t(8;21)" abnormalities were misclassified as ``inv(16)" abnormalities, and one ``t(8;21)" abnormality was misclassified as ``t(15;17)" abnormality. Overall, CS sparse 3-means has higher accuracy.

Sparse 3-means selected 50 genes (with non-zero weights). CS sparse 3-means selected 99 genes (at least one cluster pair has non-zero weight). To assess the gene selection performance by both methods, we compared the assigned gene weights results with the signature genes reported in the original paper. The original paper reported one gene signature for ``t(15;17)" abnormality: HGF, and ten gene signatures for ``t(8;21)" abnormality: LAT2, C11orf21, C15orf39, CAPG, TRH, AIF1, HSPG2, PRAME, FAM698, and LCP. The 3 clusters were identified by their greatest overlapping true subtype labels.

The gene weight assignments by both methods were summarized in Table 2 and Table 3.

\begin{table}[H]
\label{tab:1gene}
\begin{center}
	\begin{tabular}{ c | c  c  c | c }
    \hline
    Weight & inv(16) vs t(15;17) & t(15;17) vs t(8;21) & t(8;21) vs inv(16) & Sparse K-means\\ \hline 
    HGF & 0.018 & 0.148 & 0 & 0.029 \\ \hline
    \end{tabular}
    \vspace{1em}
    \caption{Weight assignment of ``t(15;17)" signature gene HGF. Clearly, HGF is discovered as a signature genes for HGF by CS sparse 3-means because it has weight for ``t(15;17)" against the other to subtypes but not for ``t(8;21)" vs ``inv(16)"}
\end{center}
\end{table}

\begin{table}[H]
\label{tab:10gene}
\begin{center}
	\begin{tabular}{  c | c  c  c | c}
    \hline
    Weight & inv(16) vs t(15;17) & t(15;17) vs t(8;21) & t(8;21) vs inv(16) & Sparse K-means\\ \hline 
    LAT2 & 0 & 0 & 0.042 & 0 \\
    C11orf21 & 0 & 0 & 0 & 0 \\
    CAPG & 0 & 0.126 & 0.030 & 0.074 \\
    TRH & 0 & 0.052 & 0.102 & 0.122 \\
    AIF1 & 0 & 0 & 0.077 & 0 \\
    HSPG2 & 0 & 0 & 0 & 0 \\
    PRAME & 0 & 0.056 & 0.168 & 0.25 \\
    C15orf39 & 0 & 0 & 0 & 0 \\
    FAM698 & 0 & 0.035 & 0 & 0 \\
    LCP & 0 & 0 & 0 & 0 \\ \hline
    \end{tabular}
    \vspace{1em}
    \caption{Weight assignment of ``t(8;21)" signature gene HGF. Clearly, most genes discovered by CS sparse 3-means indicate they were signature genes for ``t(8;21)" because their weight assignments for ``inv(16)" vs ``t(15;17)" are zero.}
\end{center}
\end{table}

Out of the 11 signature genes, sparse 3-means discovered 4 of them out of its 50 selected genes ($p < 2.5\times 10^{-6}$ by Fisher's exact test). CS sparse 3-means discovered 7 of them out if its 99 selected genes ($p < 2.5\times 10^{-10}$ by Fisher's exact test). If we take into account of the correct cluster pair specific discovery by CS sparse 3-means, the p-value is even smaller ($p < 9.0\times 10^{-13}$).

Therefore, on this leukemia gene expression dataset, CS sparse K-means achieved better clustering and gene selection accuracy than sparse K-means. Our algorithm also correctly discovered subtype specific informative genes which provide very important clues on the gene expression level alternation between specific disease subtypes. 

\section{Conclusion}

In this paper, we developed an algorithm for cluster-specific sparse clustering based on K-means. Our algorithm showed better clustering accuracy than sparse K-means on several simulated or real-world datasets. In genomics analysis, our algorithm can discover disease subtype specific informative genes by assigning weights on each gene with respect to each subtype pairs. 

In the future, we can improve the performance of our algorithm in several aspects: 1) Extend the process to Gaussian Mixtures; 2) Formally analyze convergence and correctness guarantees for this algorithm; 3) Introduce parallelism and early-stopping to speed up the process. We will also apply this algorithm to more real-world genomics dataset to assist disease subtype studies and to further develop our algorithm.

\newpage
\bibliographystyle{plain}
\bibliography{references}

\begin{thebibliography}{10}

\bibitem{appl2}
Sabrina Bouatmane, Mohamed~Ali Roula, Ahmed Bouridane, and Somaya Al-Maadeed.
\newblock Round-robin sequential forward selection algorithm for prostate
  cancer classification and diagnosis using multispectral imagery.
\newblock {\em Machine Vision and Applications}, 22(5):865--878, Sep 2011.

\bibitem{survey2014}
Girish Chandrashekar and Ferat Sahin.
\newblock A survey on feature selection methods.
\newblock {\em Computers and Electrical Engineering}, 40(1):16 -- 28, 2014.
\newblock 40th-year commemorative issue.

\bibitem{garg2016craft}
Vikas~K Garg, Cynthia Rudin, and Tommi Jaakkola.
\newblock Craft: Cluster-specific assorted feature selection.
\newblock In {\em Artificial Intelligence and Statistics}, pages 305--313,
  2016.

\bibitem{guo2010pairwise}
Jian Guo, Elizaveta Levina, George Michailidis, and Ji~Zhu.
\newblock Pairwise variable selection for high-dimensional model-based
  clustering.
\newblock {\em Biometrics}, 66(3):793--804, 2010.

\bibitem{huo2016meta}
Zhiguang Huo, Ying Ding, Silvia Liu, Steffi Oesterreich, and George Tseng.
\newblock Meta-analytic framework for sparse k-means to identify disease
  subtypes in multiple transcriptomic studies.
\newblock {\em Journal of the American Statistical Association},
  111(513):27--42, 2016.

\bibitem{appl1}
Boshu Liu, Sujun Li, Yinglin Wang, Lin Lu, Yixue Li, and Yudong Cai.
\newblock Predicting the protein sumo modification sites based on properties
  sequential forward selection (psfs).
\newblock {\em Biochemical and Biophysical Research Communications}, 358(1):136
  -- 139, 2007.

\bibitem{appl4}
Maxine Tan, Jiantao Pu, and Bin Zheng.
\newblock Optimization of breast mass classification using sequential forward
  floating selection (sffs) and a support vector machine (svm) model.
\newblock {\em International Journal of Computer Assisted Radiology and
  Surgery}, 9(6):1005--1020, Nov 2014.

\bibitem{verhaak2009prediction}
Roel~GW Verhaak, Bas~J Wouters, Claudia~AJ Erpelinck, Saman Abbas, H~Berna
  Beverloo, Sanne Lugthart, Bob L{\"o}wenberg, Ruud Delwel, and Peter~JM Valk.
\newblock Prediction of molecular subtypes in acute myeloid leukemia based on
  gene expression profiling.
\newblock {\em haematologica}, 94(1):131--134, 2009.

\bibitem{appl3}
Lili Wang, Alioune Ngom, and Luis Rueda.
\newblock Sequential forward selection approach to the non-unique
  oligonucleotide probe selection problem.
\newblock In Madhu Chetty, Alioune Ngom, and Shandar Ahmad, editors, {\em
  Pattern Recognition in Bioinformatics}, pages 262--275, Berlin, Heidelberg,
  2008. Springer Berlin Heidelberg.

\bibitem{witten2010framework}
Daniela~M Witten and Robert Tibshirani.
\newblock A framework for feature selection in clustering.
\newblock {\em Journal of the American Statistical Association},
  105(490):713--726, 2010.

\end{thebibliography}

\end{document}